\documentclass[pra,twocolumn,showpacs,superscriptaddress,
groupeaddress,preprintnumbers,amsmath,amssymb,tightenlines]{revtex4}
\usepackage{graphicx}
\newcommand{\beq}{\begin{equation}}
\newcommand{\eeq}{\end{equation}}
\newcommand{\bea}{\begin{eqnarray}}
\newcommand{\eea}{\end{eqnarray}}
\newcommand{\ba}{\begin{array}}
\newcommand{\ea}{\end{array}}
\newcommand{\bc}{\begin{center}}
\newcommand{\ec}{\end{center}}

\newcommand{\bml}{\begin{subequations}}
\newcommand{\eml}{\end{subequations}}
\newcommand{\commentout}[1]{{}}
\newcommand{\bk}{{\bf k}}

\newcommand{\C}{{\bf C}}
\newcommand{\adag}{a^\dagger}

\newcommand{\bdag}{b^\dagger}

\newcommand{\half}{\hbox{$\frac{1}{2}$}}

\newcommand{\HC}{{\rm H.c.}}
\newcommand{\eq}[1]{(\ref{#1})}
\newcommand{\etal} {{\it et al.\/}}

\newcommand{\vol}[1]{{\bf #1}}
\newcommand{\comment}[1]{{}}
\newcommand{\biN}[1]{$\left(\begin{subarray}{c} N\\ #1 \end{subarray}\,\right)$}
\newcommand{\biNma}[1]{\left(\begin{subarray}{c} N\\ #1 \end{subarray}\,\right)}

\begin{document}
\title{
Zero-Temperature Theory of Collisionless Rapid Adiabatic Passage from a\\ Fermi
Degenerate Gas of Atoms to a Bose-Einstein Condensate of Molecules}
\author{Matt Mackie}
\affiliation{QUANTOP--Danish National Research Foundation
Center for Quantum Optics, Department of Physics and Astronomy,
University of Aarhus, DK-8000 Aarhus C, Denmark}
\author{Olavi Dannenberg}
\affiliation{Helsinki Institute of Physics, PL 64, FIN-00014 Helsingin yliopisto,
Finland}
\date{\today}

\begin{abstract}
We theoretically examine a zero-temperature system of Fermi degenerate
atoms coupled to bosonic molecules via collisionless rapid adiabatic passage across a Feshbach
resonance, focusing on saturation of the molecular conversion efficiency at the
slowest magnetic-field sweep rates. Borrowing a novel {\em many-fermion} Fock-state
theory, we find that a proper model of the magnetic-field sweep can systematically remove
saturation. We also debunk the common misconception that many-body effects are responsible
for molecules existing above the two-body threshold.
\end{abstract}

\maketitle

{\em Introduction.}--Magnetoassociation creates a molecule from a pair of
colliding atoms when one of the atoms spin flips in the presence of a magnetic field
tuned near a Feshbach resonance~\cite{STW76}. Recently,
ultracold~\cite{REG03,STR03} and condensate~\cite{GRE03} molecules have
been created via magnetoassociation of a Fermi gas of atoms, in the course of efforts to
create superfluid Cooper-paired atoms~\cite{REG04,ZWI04} (see also
Refs.~\cite{CHI04}). The backbone of these experiments is rapid adiabatic passage: the ground
state of the Feshbach system is all atoms far above the molecular-dissociation threshold and all
molecules far below it, so that a slow sweep of the magnetic field from one extreme to the other
converts atoms into diatomic molecules.

\begin{figure}[t]
\centering
\includegraphics[width=8.0cm]{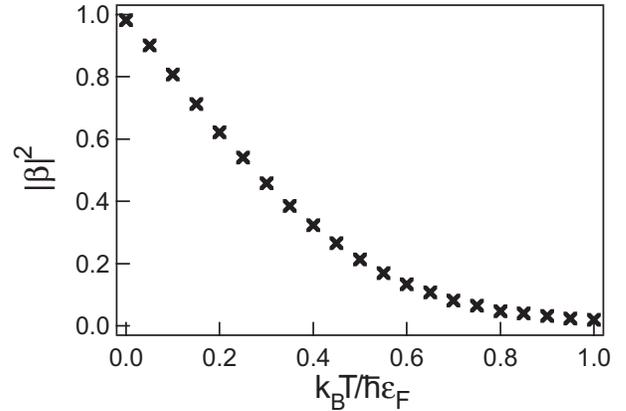}
\caption{Predicted temperature dependence for the efficiency ($|\beta|^2$) of collisionless rapid
adiabatic passage from quantum degenerate $^{40}$K atoms to $^{40}\rm{K}_2$ molecules.
As the temperature in Fermi units ($\hbar\varepsilon_F/k_B$) is lowered, atoms become more
likely to be affected by the dynamical stability that forms molecules. The magnetic field was
swept linearly at the (inverse) rate
$1/\dot{B}=400\mu\rm{s}/\rm{G}$. Figure reproduced from Ref.~\cite{JAV04}.}
\label{T_PRED}
\end{figure}

Finite-temperature mean-field theory of magnetoassociation of a Fermi gas of atoms leads
to two types of instabilities against molecule formation. One is the thermodynamic
instability of a Fermi sea against the formation of Cooper pairs~\cite{TIN75}, a trait of
superconductors whose analog is passed on to Feshbach-resonant superfluids~\cite{TIM01}.
A thermodynamical instability occurs because pairing lowers the energy, and coupling
to a reservoir with a low enough temperature leaves the system prone to pairing. The
other is a dynamical instability, whereby the larger state space of the molecules, owing
somewhat to Pauli blocking, leaves the atoms prone to spontaneous
association~\cite{JAV04}. The role that temperature plays in this process is an open
question experimentally, as well as a matter of theoretical contention.

Physically~\cite{JAV04}, high temperature lessens the chance of
an atom occupying an arbitrary level in the Fermi sea, the dynamical
instability becomes less effective and the efficiency of even the slowest rapid adiabatic passage
therefore saturates (c.f., Fig.~\ref{T_PRED}). The mean-field theory
behind this understanding agrees semi-quantitatively with experiments~\cite{REG03};
nevertheless, a recent zero-temperature Landau-Zener theory predicts that saturation is
fundamental to the collisionless regime~\cite{CHW04}. If temperature is not a limiting factor,
then any zero-temperature model of collisionless rapid adiabatic passage should ultimately
display saturation, e.g., a Fock-state approach similar to the theory of cooperative association of
Bose-Fermi mixtures of atoms into Fermi molecules~\cite{DAN03}. Unfortunately, computing
power is presently sufficient for calculations with only about 20 atoms total at best, precluding
any brute-force~\cite{DAN03} test of saturation. Here we apply a novel
{\em large-fermion-number} theory~\cite{TIK04} to demonstrate near-unit-efficient
collisionless rapid adiabatic passage in the limit of zero temperature, thereby ruling out any
fundamental ceiling to the molecular conversion, and bolstering Ref.~\cite{JAV04} (and also
Ref.~\cite{PAZ04}).

This development is outlined as follows. After briefly introducing the collisionless
model, we focus on rapid adiabatic passage and confirm the reduced-space
mapping~\cite{TIK04} by comparison with exact few-particle results.
Increasing the total particle number to $2\times10^2$, we then observe what appears to be
saturation at about $\sim50\%$. However, including fluctuation effects in the
rate at which the system is swept across the Feshbach resonance, we find that saturation can be
systematically removed, and near-unit efficiency can be achieved for any particle number.
Lastly, from the single pair results we also debunk the commonly held notion that many-body
effects are responsible for the existence of molecules above the threshold for molecular
dissociation.

{\em Collisionless Gas Model.}--We model an ideal two-component gas of fermionic
atoms coupled by a Feshbach resonance to bosonic molecules. In the language of
second-quantization, an atom of mass $m$ and momentum $\hbar\bk$ is described by the
annihilation operator $a_{\bk,1(2)}$, and a molecule of mass $2m$ and similar momentum is
described by the annihilation operator $b_\bk$. All operators obey their (anti)commutation
relations. The microscopic Hamiltonian for such a freely-ideal system is written
\bea
\frac{H}{\hbar} &=&
\sum_{\bk} \left[\left(\epsilon_k-\mu\right)
  \adag_{\bk,\sigma}a_{\bk,\sigma}+(\half\epsilon_k+\delta-\mu_{\rm mol})\bdag_\bk
b_\bk\right]
\nonumber\\&&
+\kappa\sum_{\bk,\bk'}\left[\bdag_{\bk+\bk'}a_{\bk,1}a_{\bk',2}+\HC\right],
\label{MICRO_HAM}
\eea
where repeated greek indices imply a summation ($\sigma=1,2$). The free-particle energy
is $\hbar\epsilon_{k}=\hbar^2 k^2/2m$, the atom (molecule) chemical potential is
$\hbar\mu_{\sigma(\rm mol)}$, and the detuning $\delta$ is a measure of the
binding energy of the molecule ($\delta>0$ is taken as above threshold), the
mode-independent atom-molecule coupling is
$\kappa\propto1/\sqrt{V}$ with $V$ is the quantization volume.

We have already imposed the ideal conditions
for atom-molecule conversion with $\mu_1=\mu_2=\mu$. An appropriate unitary
transformation then shuffles $\mu$ into the definition of $\mu_{\rm mol}$ which, in
turn, can be absorbed into the detuning and written off as an
effectively dc bias (see also Refs.~\cite{DAN03}). Since magnetoassociation usually
occurs much faster than any trapped-particle motion,  an explicit trap can be neglected along
with the free-particle energies $\epsilon_k$~\cite{CRUTCH_FOOT}. For the sake of
simplicity, and to compare with Ref.~\cite{JAV04}, we neglect all molecular modes except the
$\bk+\bk'=0$ mode,
$b_0\equiv b$, so that
\bea
\frac{H}{\hbar} &=&
\delta\bdag b
+\kappa\sum_\bk\left[\bdag a_{\bk,1}a_{-\bk,2}+\HC\right].
\label{EFF_HAM}
\eea

 Absent losses, the total particle number is conserved,
$2\langle\bdag b\rangle+\sum_\bk\langle\adag_{\bk\sigma} a_{\bk\sigma}\rangle
=2n+\sum_{\bk,\sigma} n_{\bk\sigma}=2N$,
where $n$ is the number of molecules, $n_{k,\sigma}=0,1$ is the number of
atoms per mode ($k$) per species ($\sigma$), and  $2N$ is the total number of atoms were all
the molecules to dissociate. For a fixed number of particles equal to the number of fermion
modes, the Fock-state wavefunction is~\cite{TIK04}
\beq
|\psi(t)\rangle=\sum_{m'=0}^N\sum_{\{n_k\}}C_{N-m',n_1,\ldots,n_N}(t).
  \left| N-m',n_1,\ldots,n_N\right\rangle
\eeq
The time dependence of the system is determined by the Schr\"odinger equation,
$i\hbar\partial_t|\psi\rangle=H|\psi\rangle$, so that the Hamiltonian~\eq{EFF_HAM}
yields~\cite{TIK04}
\bea
i\dot\C_m&=&[N-m]\delta\C_m
\nonumber\\&&
  +\kappa\left[\sqrt{N-m+1}\,D_m^{m-1}\C_{m-1}
\right.\nonumber\\&&\left.\hspace{0.75cm}
  +\sqrt{N-m}\,D_m^{m+1}\C_{m+1}\right].
\label{FULL_EQM}
\eea
Here $\C_m(t)\equiv\C_{N-m,n_1,\ldots,n_N}(t)$ is a column vector of all the amplitudes
corresponding to the \biN{m} possible arrangements of $m$ atom pairs among
the $N$ available fermion modes, and $D_I^J$ is an \biN{I}$\times$\biN{J}
dimensional matrix that contains only unit and zero elements determined
by $\C_I$ and $\C_J$. The problem with the system~\eq{FULL_EQM} is that there are
$2^N$ amplitudes, which limits most numerical experiments in rapid adiabatic passage to
about $N=10$ (see also Refs.~\cite{DAN03});  however, by multiplying
Eqs.~\eq{FULL_EQM} by the appropriate column vector ${\rm\bf u}_{m,N}$, any
redundant amplitudes can be eliminated~\cite{TIK04}. The remaining $N+1$ amplitudes
evolve in time according to~\cite{TIK04}
\bea
i\dot\alpha_m&=&
  [N-m]\delta\alpha_m
\nonumber\\&&
  +\kappa\left[\sqrt{m}\,(N-m+1)\alpha_{m-1}
\right.\nonumber\\&&\left.\hspace{0.75cm}
  +\sqrt{m+1}\,(N-m)\alpha_{m+1}\right],
\label{RED_EQM}
\eea
where the sum of all \biN{m} amplitudes with $N-m$ molecules and $m$ free
atom pairs is defined as
$\sqrt{\biNma{m}}\,\alpha_m\equiv{\rm\bf
u}_{m,N}\C_m=\sum_{n_k}C_{N-m,n_1,\ldots,n_N}$
(with $\alpha_m$ normalized to the number of permutations of $m$ atoms in $N$ states).
Lastly we will need the molecular fraction 
$|\beta|^2=2\langle\bdag b\rangle/(2N)=(1/N)\sum_{m=0}^N(N-m)|\alpha_m|^2$.

\begin{figure}
\centering
\includegraphics[width=8.0cm]{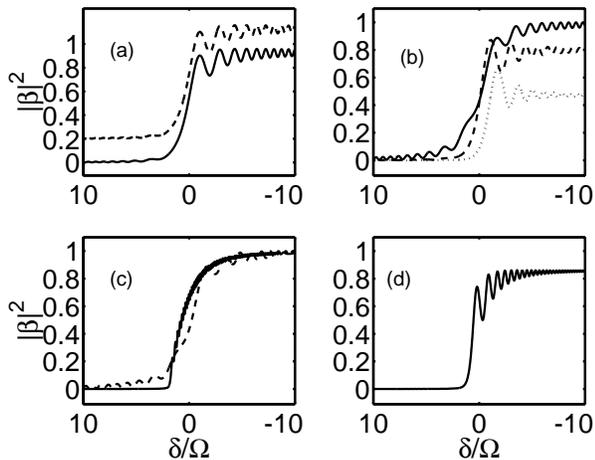}
\caption{Molecular condensate fraction as a function of the
detuning in rapid adiabatic passage across a Feshbach resonance, beginning above threshold
($\delta>0$). (a)  For
$N=4$, comparison of solution to the full equations of motion~[Eq.~\eq{FULL_EQM},
dashed line] with the solution to the reduced-space equations of motion~[Eq.~\eq{RED_EQM},
solid line]. The full results are shifted for clarity; the two
calculations are otherwise indistinguishable. The detuning sweep model is
$\delta(t)=-\xi\Omega^2t$, with
$\xi=1$. (b) Using the same sweep model, we find apparent saturation for increasing particle
number:
$N=4$ (solid line), $N=10$ (dashed line), and $N=10^2$ (dotted line).
(c) A fluctuation-adjusted sweep model, $\delta(t)=-\xi(\Omega/\ln{N})^2t$, leads to near-unit
efficiency for $N=10^2$ (solid line), as well as solid agreement with the $N=1$ results. Here
the dimensionless sweep rate is again unity, $\xi=1$. (d) Results for $N=10^2$ and $\xi=7.6$,
an estimate of a zero-temperature sweep for $^{40}$K (see text).}
\label{ALL}
\end{figure}

{\em Rapid Adiabatic Passage.}--Putting fluctuations~\cite{VAR01} momentarily
aside, the relevant frequency scale is
$\Omega=\sqrt{N}\kappa\propto\sqrt{\rho}$~\cite{JAV99,DAN03}, where
$\sqrt{\rho}$ is the so-called collective enhancement factor. ``Adiabatic "is therefore defined
qualitatively as the detuning changing by an amount $\Omega$ in a time $1/\Omega$, or
$|\dot\delta|\alt\Omega^2$. Modeling the time dependent detuning as $\delta=-\xi\Omega^2t$,
sweeps with $\xi\sim1$ should qualify as adiabatic. Off hand, Fig.~\ref{ALL}(a) confirms this
intuition for $N=4$; also, noting that the full results are shifted for clarity, the
reduced system~\eq{RED_EQM} indeed reproduces the full
system~\eq{FULL_EQM}.  Making a more full use of the reduced-space theory,
Fig.~\ref{ALL}(b) illustrates that the efficiency of rapid adiabatic passage in fact decreases for
increasing particle number, saturating at about 50\% for
$N=10^2$. Nevertheless, if we account for fluctuations, then the relevant frequency scale is
$\Omega/\ln{N}$~\cite{VAR01}. Now the detuning should change by
$\Omega/\ln{N}$ in a time $(\Omega/\ln{N})^{-1}$, suggesting the detuning-sweep model
$\delta(t)=-\xi(\Omega/\ln{N})^2t$. Indeed, Fig.~\ref{ALL}(c) shows that the $N=10^2$ and
$N=1$ results agree nicely, and are absent any evident saturation. 

We can also make a rough comparison with the zero-temperature limit in Fig.~\ref{T_PRED}.
Magnetic fields are converted into detunings according to
$\delta=\Delta_\mu(B-B_0)/\hbar$, where the difference in magnetic moments
between the atom pair and a molecule is
$\Delta_\mu$, and $B_0$ is the magnetic-field position of resonance. For
$N=10^5$ atoms of $^{40}$K in a typical~\cite{REG03} trap the peak density
$\rho=2\times10^{13}\rm{cm}^{-3}$, so that the coupling strength is
$\Omega=0.3\times2\pi$~MHz~\cite{JAV04}; the difference in magnetic moments is
$\Delta_\mu\approx0.19\mu_0$~\cite{JAV04}, where $\mu_0$ is the Bohr magneton.
The results of  Fig.~\ref{T_PRED} are for
$1/\dot{B}=400\mu\rm{s}/\rm{G}$~\cite{JAV04}, which corresponds to
$\xi=(\ln{N})^2\Delta_\mu\dot{B}/(\hbar\Omega^2)\approx7.9$ for $N=10^5$ atoms per
species. Of course, even the reduced-space model~\cite{TIK04} cannot handle $N=10^5$
atoms, but for $\xi=7.9$ then $N=10^2$ will actually {\em underestimate} the $N=10^5$
results. Hence the already good agreement between Fig.~\ref{ALL}(d) and 
Fig.~\ref{T_PRED} would actually improve if resources were available to manage the correct
number of particles.

We pause briefly to justify the ideal gas model. The collisional interaction
strength is roughly $\Lambda=2\pi\hbar\rho a/m$, where $a$ is the off-resonant atomic
$s$-wave scattering length. The $^{40}$K scattering length is $a=176a_0$~\cite{BOH00},
with $a_0$ the Bohr radius. For a typical density
$\rho\sim10^{13}\rm{cm}^{-3}$, the collisional coupling strength, in units of the
atom-molecule coupling, is
$|\Lambda|/\Omega\approx10^{-3}$. Collisions should therefore be broadly negligible. In
particular, a system of Fermi atoms coupled to Bose molecules is formally identical to a system
of only  bosons~\cite{TIK04}, and collisions are negligible for bosons
under such conditions~\cite{ISH04}.

Also, it should be noted that, because we have chosen $\Omega$ as the frequency scale, the
above results are actually for a resonance (atom-molecule coupling) of arbitrary strength.
However, the model~\eq{RED_EQM} is broadly equivalent to the two-mode model in coherent
association of condensate~\cite{TIK04}, and it is well known that strong coupling can lead to
dissociation to modes lying outside the two-mode system~\cite{JAV99,MAC02}, so-called
rogue dissociation. Nevertheless, if the sweep is directed from above to below threshold, then
rogue dissociation is negligible and the two-mode model is a good approximation. Hence, the
above results are expectedly reasonable to describe a sweep across an arbitrarily strong
resonance.

\begin{figure}[b]
\centering
\includegraphics[width=8.0cm]{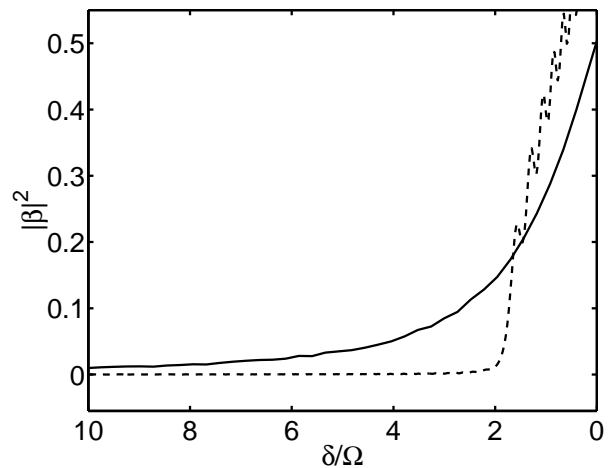}
\caption{Fraction of above-threshold molecular condensate for $N=100$ (from
Fig.~\ref{ALL}(c), dashed) compared to ground-state $N=1$ (solid line). Evidently,
above-threshold molecules exist in the absence of many-body effects and, oddly enough, such
effects can act to {\em suppress} the molecular fraction for positive detunings
$\delta/\Omega\agt2$.}
\label{ABV_THRSH}
\end{figure}

Before closing, we turn to a related matter of principle: the nature of above-threshold
molecules.  Below threshold ($\delta<0$), Fourier analysis delivers the binding energy,
$\hbar\omega_B<0$, of the Bose-condensed molecules~\cite{MAC02,JAV04}:
$\omega_B-\delta-\Sigma'(\omega_B)+i\eta=0,$
where $\Sigma'(\omega_B)$ is the finite self-energy of the Bose molecules and
$\eta=0^+$. Tuning the system above the two-body threshold ($\delta>0$) gives an imaginary
$\omega_B$, and the bound state ceases to exist; nevertheless,
Fig.~\ref{ABV_THRSH} shows a large $N=1$ molecular fraction. This result is not really a
surprise, since the fraction of molecules must vary continuously
from zero to unity across threshold. We conclude that any theory in which molecules
abruptly cease to exist at threshold, while useful in their own right (e.g., for modeling binding
energies~\cite{JAV04}), are not a good rule of thumb for predicting the existence of
above-threshold molecules.  Our interpretation is that, as usual in cooperative behavior, a
macroscopic number of particles respond as a unit to a given external drive, thereby mimicking
one- or two-body physics. At the least, this implies that many-body effects are sufficient
but not necessary for the existence of above-threshold molecules. Moreover, we see in
Fig.~\ref{ABV_THRSH} that the above-threshold molecular fraction for $\delta/\Omega\alt2$
is actually suppressed for the many-body case $N=10^2$.

Of course, the idea of many-body stabilization of above-threshold molecules generally arises in
the context of equilibrium thermodynamics, whereas the collisionless model
describes non-equilibrium processes. The many-body suppression of the above-threshold
molecular fraction may or may not carry over to the collisional regime (although we find
elsewhere that, to a certain degree, it may~\cite{MAC04}). However, we expect that the
two-body equilibration time is sufficiently long compared to the atom-molecule conversion
timescale that, even in the presence of collisions, a two-body system can always be considered
out of equilibrium; hence, the two-body ground state of Fig.~\ref{ABV_THRSH} should apply
to the collisional regime as well.

{\em Conclusions.}--We have investigated saturation in collisionless rapid adiabatic passage
from a two-component degenerate Fermi gas to a Bose-Einstein condensate of molecules.
Saturation indeed arises, but can be systematically eliminated by introducing the
timescale appropriate to cooperative interference effects. Physically, cooperative interference
effects arise from adding up the various pathways coupling the states having $N-m$ molecules
and $m$ dissociated atom pairs with the states having one more (less) molecules and one less
(more) dissociated pair, and the timescale for $N$-particle interference turns out to be
$\sim\ln{N}/\Omega$~\cite{TIK04}. It then makes perfect sense that cooperative
(near-unit-efficient and macroscopic) rapid adiabatic passage will only occur over a timescale
that is commensurate with constructive interference.  Next we saw that our zero-temperature
model agrees semi-quantitatively with our mean-field model~\cite{JAV04}, indicating that
temperature~\cite{JAV04} and pair correlations~\cite{PAZ04} are--as of yet--the main
obstacles to collisionless cooperative conversion to molecules with near-unit efficiency. 
Finally, whereas studies of Feshbach resonances for both fermions and bosons have implicated
many-body effects in the existence of molecules above the two-body threshold for dissociation,
we find that it is not necessary to invoke many-body effects to explain the existence of
above-threshold molecules.

{\em Acknowledgements.}--One of us (O.D.) kindly thanks the Magnus Ehrnrooth Foundation
for support.


\begin{thebibliography}{99}
\bibitem{CRUTCH_FOOT}
  This means we will be able to fully account for the atoms being fermions without, per say,
stacking them in trap levels, and that densities are understood to be uniform.
\end{thebibliography}
\end{document}